\documentclass[prd,onecolumn,nofootinbib,showpacs]{revtex4}
\usepackage{amssymb,amsmath,bm,graphicx}

\begin{document}

\title{NONSINGULAR, BIG-BOUNCE COSMOLOGY FROM SPINOR-TORSION COUPLING}

\author{{\bf Nikodem Pop{\l}awski}}

\affiliation{Department of Physics, Indiana University, Bloomington, Indiana, USA}
\email{nikodem.poplawski@gmail.com}

\noindent
{\em Physical Review D}\\
Vol. {\bf 85}, No. 10 (2012) 107502\\
\copyright\,American Physical Society
\vspace{0.4in}

\begin{abstract}
The Einstein-Cartan-Sciama-Kibble theory of gravity removes the constraint of general relativity that the affine connection be symmetric by regarding its antisymmetric part, the torsion tensor, as a dynamical variable.
The minimal coupling between the torsion tensor and Dirac spinors generates a spin-spin interaction which is significant in fermionic matter at extremely high densities.
We show that such an interaction averts the unphysical big-bang singularity, replacing it with a cusp-like bounce at a finite minimum scale factor, before which the Universe was contracting.
This scenario also explains why the present Universe at largest scales appears spatially flat, homogeneous and isotropic.
\end{abstract}


\maketitle

The Einstein-Cartan-Sciama-Kibble (ECSK) theory of gravity, like general relativity (GR), is based on the gravitational Lagrangian density that is proportional to the curvature scalar $R$ \cite{LL}.
It removes, however, the GR constraint that the affine connection $\Gamma^{\,\,k}_{i\,j}$ be symmetric by regarding the antisymmetric part of the connection, the torsion tensor $S^k_{\phantom{k}ij}=\Gamma^{\,\,\,k}_{[i\,j]}$, as a dynamical variable \cite{KS}.
Varying the total Lagrangian density $-\frac{1}{2\kappa}R\sqrt{-g}+\mathfrak{L}_\textrm{m}$, where $\mathfrak{L}_\textrm{m}$ is the Lagrangian density of matter, with respect to the contortion tensor $C_{ijk}=S_{ijk}+S_{jki}+S_{kji}$ gives the Cartan equations
\begin{equation}
S^j_{\phantom{j}ik}-S_i \delta^j_k+S_k \delta^j_i=-\frac{1}{2}\kappa s^{\phantom{ik}j}_{ik},
\label{Cartan}
\end{equation}
where $S_i=S^k_{\phantom{k}ik}$ and $s^{ijk}=2(\delta\mathfrak{L}_\textrm{m}/\delta C_{ijk})/\sqrt{-g}$ is the spin tensor.
These equations are linear and algebraical: torsion is proportional to spin density and vanishes outside material bodies.

Varying the total Lagrangian density with respect to the metric tensor $g_{ik}$ gives the Einstein equations with terms on the curvature side that are quadratic in the torsion tensor.
Substituting (\ref{Cartan}) into these equations leads to the Einstein-Cartan equations $G_{ik}=\kappa(T_{ik}+U_{ik})$, where $G_{ik}$ is the Einstein tensor, $T_{ik}=2(\delta\mathfrak{L}_\textrm{m}/\delta g^{ik})/\sqrt{-g}$ is the energy-momentum tensor, and
\begin{equation}
U^{ik}=\kappa\biggl(-s^{ij}_{\phantom{ij}[l}s^{kl}_{\phantom{kl}j]}-\frac{1}{2}s^{ijl}s^k_{\phantom{k}jl}+\frac{1}{4}s^{jli}s_{jl}^{\phantom{jl}k}+\frac{1}{8}g^{ik}(-4s^l_{\phantom{l}j[m}s^{jm}_{\phantom{jm}l]}+s^{jlm}s_{jlm})\biggr)
\label{Einstein}
\end{equation}
is the correction to the energy-momentum tensor generated by torsion and quadratic in the spin tensor \cite{Hehl}.
The spin tensor also appears in $T_{ik}$ because $\mathfrak{L}_\textrm{m}$ depends on torsion.
The corrections from the spin tensor to the right-hand side of the Einstein equations are significant only at extremely high densities, on the order of the Cartan density \cite{non}.
Below this density, the predictions of the ECSK theory do not differ from the predictions of GR.
In vacuum, where torsion vanishes, this theory reduces to GR.
The ECSK gravity therefore passes all observational and experimental tests of GR \cite{Hehl}.

In the ECSK theory, the Dirac Lagrangian density for a free spinor $\psi$ with mass $m$, minimally coupled to the gravitational field, is given by $\mathfrak{L}_\textrm{m}=\frac{i}{2}\sqrt{-g}(\bar{\psi}\gamma^i\psi_{;i}-\bar{\psi}_{;i}\gamma^i\psi)-m\sqrt{-g}\bar{\psi}\psi$, where $g$ is the determinant of the metric tensor and $\gamma^i$ are the Dirac matrices obeying $\gamma^{(i}\gamma^{k)}=g^{ik}I$.
Semicolon denotes a covariant derivative with respect to the affine connection \cite{KS,Hehl}:
\begin{equation}
\psi_{;k}=\psi_{:k}+\frac{1}{4}C_{ijk}\gamma^{[i}\gamma^{j]}\psi,\,\,\,\bar{\psi}_{;k}=\bar{\psi}_{:k}-\frac{1}{4}C_{ijk}\bar{\psi}\gamma^{[i}\gamma^{j]},
\label{covariant}
\end{equation}
and colon denotes a Riemannian covariant derivative with respect to the Christoffel symbols.
We use the units in which $c=\hbar=k_\textrm{B}=1$, so $\kappa=8\pi G=m^{-2}_{\textrm{P}}$, where $m_{\textrm{P}}$ is the reduced Planck mass.
For a Dirac field, the spin tensor is completely antisymmetric:
\begin{equation}
s^{ijk}=-e^{ijkl}s_l,\,\,\,s^i=\frac{1}{2}\bar{\psi}\gamma^i\gamma^5\psi,
\label{spin}
\end{equation}
where $e^{ijkl}=\epsilon^{ijkl}/\sqrt{-g}$, $\epsilon^{ijkl}$ is the Levi-Civita permutation symbol, and $s^i$ is the Dirac spin pseudovector.
The Cartan equations for such a field give therefore the completely antisymmetric contortion tensor \cite{KS}:
\begin{equation}
C_{ijk}=S_{ijk}=\frac{1}{2}\kappa e_{ijkl}s^l.
\label{torsion}
\end{equation}
Substituting (\ref{spin}) into (\ref{Einstein}) gives
\begin{equation}
U^{ik}=\frac{1}{4}\kappa(2s^i s^k+s^l s_l g^{ik}).
\label{first}
\end{equation}

Varying $\mathfrak{L}_\textrm{m}$ with respect to the spinor adjoint conjugate $\bar{\psi}$ gives the Dirac equation $i\gamma^k\psi_{;k}=m\psi$, which can be written as $i\gamma^k\psi_{:k}=m\psi-\frac{3}{8}\kappa(\bar{\psi}\gamma^k\gamma^5\psi)\gamma_k\gamma^5\psi$ and whose conjugate is $-i\bar{\psi}_{:k}\gamma^k=m\bar{\psi}-\frac{3}{8}\kappa(\bar{\psi}\gamma^k\gamma^5\psi)\bar{\psi}\gamma_k\gamma^5$ \cite{Hehl,HD}.
Putting this equation in the energy-momentum tensor corresponding to the Dirac Lagrangian, $T_{ik}=\frac{i}{2}(\bar{\psi}\delta^j_{(i}\gamma_{k)}\psi_{;j}-\bar{\psi}_{;j}\delta^j_{(i}\gamma_{k)}\psi)-\frac{i}{2}(\bar{\psi}\gamma^j\psi_{;j}-\bar{\psi}_{;j}\gamma^j\psi)g_{ik}+m\bar{\psi}\psi g_{ik}$, gives
\begin{equation}
T_{ik}=\frac{i}{2}(\bar{\psi}\delta^j_{(i}\gamma_{k)}\psi_{;j}-\bar{\psi}_{;j}\delta^j_{(i}\gamma_{k)}\psi).
\label{energymomentum}
\end{equation}
Substituting (\ref{covariant}) and (\ref{torsion}) into (\ref{energymomentum}) gives
\begin{equation}
T_{ik}=\frac{i}{2}(\bar{\psi}\delta^j_{(i}\gamma_{k)}\psi_{:j}-\bar{\psi}_{:j}\delta^j_{(i}\gamma_{k)}\psi)+\frac{1}{2}\kappa(-s_i s_k+s^l s_l g_{ik}).
\label{second}
\end{equation}
The combined energy-momentum tensor for a Dirac field on the right-hand side of the Einstein-Cartan equations is thus
\begin{equation}
T_{ik}+U_{ik}=\frac{i}{2}(\bar{\psi}\delta^j_{(i}\gamma_{k)}\psi_{:j}-\bar{\psi}_{:j}\delta^j_{(i}\gamma_{k)}\psi)+\frac{3}{4}\kappa s^l s_l g_{ik},
\label{combined}
\end{equation}
which has been found by Kerlick \cite{Ker}.
This expression agrees with \cite{dark}, where we derived it from the Hehl-Datta equation, which is the Dirac equation upon substituting (\ref{covariant}), (\ref{spin}) and (\ref{torsion}) \cite{HD}.
The first term on the right of (\ref{combined}) is the GR part of the energy-momentum tensor for a Dirac field and can be macroscopically averaged at cosmological scales as a perfect fluid with the energy density $\epsilon$ and pressure $p$.
In the comoving frame of reference, in which $g_{0\alpha}=0$ ($\alpha$ denotes space indices) and the four-velocity $u^i$ of the cosmological fluid satisfies $u^0=1$ and $u^\alpha=0$ \cite{Ker}, the relation $s^i u_i=0$ gives $s^0=0$.
Accordingly, the second term on the right of (\ref{combined}) is equal to $-\frac{3}{4}\kappa{\bf s}^2 g_{ik}$, where ${\bf s}$ is the spatial spin pseudovector which measures the spatial density of spin.
The average value of its square is $\langle{\bf s}^2\rangle=\frac{3}{4}n^2$, where $n$ is the fermion number density.
The averaged second term on the right of (\ref{combined}) acts thus like a perfect fluid with a negative energy density:
\begin{equation}
\tilde{\epsilon}=-\tilde{p}=-\alpha n^2,\,\,\,\alpha=\frac{9}{16}\kappa.
\label{effective}
\end{equation}

Hehl, von der Heyde, and Kerlick have used the spin-fluid approximation of fermionic matter, $s_{ijk}=s_{ij}u_k$ and $s_{ij}u^j=0$, to show that the spin-density contribution to $T_{ik}+U_{ik}$ behaves like a stiff matter with $\tilde{\epsilon}=\tilde{p}=-\frac{1}{4}\kappa s^2$, where $s^2=\frac{1}{2}s_{ik}s^{ik}=\frac{1}{8}n^2$ \cite{avert,average}.
This behavior is significant in spin fluids at extremely high densities, even without spin polarization, leading to gravitational repulsion and avoidance of curvature singularities by violating the energy condition of the singularity theorems \cite{avert}.
Trautman, Kuchowicz, and others have shown that such a repulsion replaces the big-bang singularity with a nonsingular big bounce, before which the Universe was contracting \cite{reg,Kuch}.
In contrast to spin fluids, Dirac spinors coupled to torsion enhance the energy condition for the formation of singularities \cite{Ker,OC}.

The spin-fluid model can be derived as the particle approximation of multiple expansion of the integrated conservation laws in the ECSK gravity \cite{NSH}.
The particle approximation for Dirac fields, however, is not self-consistent \cite{non}.
The spin-fluid description also violates the cosmological principle \cite{cosmo1}.
In this paper, we use the Dirac form of the spin tensor for fermionic matter, $s_{ijk}=s_{[ijk]}$ \cite{KS,Ker,OC}, which follows directly from the Dirac Lagrangian and is consistent with the cosmological principle \cite{cosmo2}.
We show that the minimal coupling between the torsion tensor and Dirac fermions, despite enhancing the energy condition, also averts the big-bang singularity.

As in \cite{infl}, we consider a closed, homogeneous and isotropic universe, described by the Friedman-Lema\^{i}tre-Robertson-Walker (FLRW) metric.
In the isotropic spherical coordinates, this metric is given by $ds^2=dt^2-a^2(t)(1+kr^2/4)^{-2}(dr^2+r^2 d\vartheta^2+r^2\mbox{sin}^2\vartheta d\varphi^2)$, where $a(t)$ is the scale factor and $k=1$.
The corresponding Einstein equations for the combined energy-momentum tensor (\ref{combined}) in the comoving frame become the Friedman equations (the cosmological constant is negligible in the early Universe):
\begin{eqnarray}
& & {\dot{a}}^2+k=\frac{1}{3}\kappa\Bigl(\epsilon-\alpha n^2\Bigr)a^2, \label{Fri1} \\
& & {\dot{a}}^2+2a\ddot{a}+k=-\kappa\Bigl(p+\alpha n^2\Bigr)a^2,
\label{Fri2}
\end{eqnarray}
where dot denotes differentiation with respect to the cosmic time $t$.
These equations yield the conservation law $\frac{d}{dt}\bigl((\epsilon-\alpha n^2)a^3\bigr)+(p+\alpha n^2)\frac{d}{dt}(a^3)=0$, which gives
\begin{equation}
a^3 d\epsilon-2\alpha a^3 ndn+(\epsilon+p)d(a^3)=0.
\label{law}
\end{equation}

As in \cite{th}, we use $\epsilon$, $p$ and $n$ for ultrarelativistic matter in kinetic equilibrium: $\epsilon(T)=\frac{\pi^2}{30}g_\star(T)T^4$, $p(T)=\frac{\epsilon(T)}{3}$ and $n(T)=\frac{\zeta(3)}{\pi^2}g_n(T)T^3$, where $T$ is the temperature of the early Universe \cite{cosmo}.
The effective numbers of thermal degrees of freedom are $g_\star(T)=g_\textrm{b}(T)+\frac{7}{8}g_\textrm{f}(T)$ and $g_n(T)=\frac{3}{4}g_\textrm{f}(T)$ (only fermions contribute to torsion), where $g_\textrm{b}=\sum_i g_i$ is summed over relativistic bosons, $g_\textrm{f}=\sum_i g_i$ is summed over relativistic fermions, and $g_i$ is the number of the spin states for each particle species $i$.
Substituting these values to (\ref{law}) gives
\begin{equation}
\frac{dT}{T}-\frac{3\alpha h^2_n}{2h_\star}TdT+\frac{da}{a}=0,
\label{diff}
\end{equation}
where $h_\star=\frac{\pi^2}{30}g_\star(T)$ and $h_n=\frac{\zeta(3)}{\pi^2}g_n(T)$ can be assumed constant in the range of $T$ considered.\footnote{
For constant values of $g_\star$ and $g_n$, the relations $\epsilon\propto T^4$, $p=\frac{\epsilon}{3}$ and $n\propto T^3$ are consistent with a relation $\frac{dn}{n}=\frac{d\epsilon}{\epsilon+p}$ used in \cite{infl}.
If $\tilde{\epsilon}=\tilde{p}\propto n^2$ as in \cite{avert,average,infl}, then we also have $\frac{dn}{n}=\frac{d(\epsilon+\tilde{\epsilon})}{\epsilon+\tilde{\epsilon}+p+\tilde{p}}$.
}
Integrating (\ref{diff}) gives
\begin{equation}
a=\frac{a_\textrm{r} T_\textrm{r}}{T}\mbox{exp}\biggl(\frac{3\alpha h^2_n}{4 h_\star}T^2\biggr),
\label{int}
\end{equation}
where $a_\textrm{r}$ is the scale factor at a reference temperature $T_\textrm{r}$.

The function $a(T)$ (\ref{int}) is not monotonic.
As $T$ increases, $a$ decreases until $T$ reaches a critical temperature $T_\textrm{cr}$ given by $\frac{da}{dT}(T_\textrm{cr})=0$,
\begin{equation}
T_\textrm{cr}=\biggl(\frac{2h_\star}{3\alpha h^2_n}\biggr)^{1/2},
\label{critical}
\end{equation}
and then increases.
Since an increasing function $a(T)$ is unphysical, $a_\textrm{cr}=a(T_\textrm{cr})>0$ is the smallest allowed value of the scale factor:
\begin{equation}
a_\textrm{cr}=a_\textrm{r} T_\textrm{r}\biggl(\frac{3e\alpha h^2_n}{2h_\star}\biggr)^{1/2}.
\label{minimum}
\end{equation}
The Universe is therefore nonsingular: $a\ge a_\textrm{cr}$.
For $T\ll T_\textrm{cr}$, (\ref{int}) reduces to $a=\frac{a_\textrm{r} T_\textrm{r}}{T}$, which is satisfied in the radiation-dominated era.

To verify that $a_\textrm{cr}$ is the minimum scale factor of the Universe, we substitute (\ref{diff}) into (\ref{Fri1}) without the negligible term $k=1$, obtaining
\begin{equation}
\dot{T}^2\biggl(\frac{1}{T^2}-\frac{3\alpha h^2_n}{2h_\star}\biggr)^2=\frac{\kappa}{3}(h_\star T^2-\alpha h^2_n T^4).
\label{scaling}
\end{equation}
Denoting $\beta=T^{-1}$ and $\beta_\textrm{cr}=T^{-1}_\textrm{cr}$ leads, with (\ref{critical}), to
\begin{equation}
|\dot{\beta}|=\sqrt{\frac{\kappa h_\star}{3}}\frac{\sqrt{\beta^2-\frac{2}{3}\beta^2_\textrm{cr}}}{\beta^2-\beta^2_\textrm{cr}},
\label{dynamics1}
\end{equation}
which yields $\beta\ge\beta_\textrm{cr}$ and $T\le T_\textrm{cr}$.
Equation (\ref{int}) gives then $a\ge a_\textrm{cr}$.
We make a substitution:
\begin{equation}
\beta=\sqrt{\frac{2}{3}}\beta_\textrm{cr}\mbox{cosh}\eta,
\label{param1}
\end{equation}
where $\eta$ is a parameter satisfying $|\eta|\ge\eta_\textrm{cr}=\mbox{arcosh}\sqrt{\frac{3}{2}}$.
Putting (\ref{param1}) in (\ref{dynamics1}) gives $t_0(\frac{2}{3}\mbox{cosh}^2\eta-1)|\frac{d\eta}{dt}|=1$, where $t_0=\beta^2_\textrm{cr}\sqrt{\frac{3}{\kappa h_\ast}}$ is a characteristic time scale of the torsion-dominated era.
Integrating this equation, with $t=0$ taken as the instant at which $\eta$ jumps from $-\eta_\textrm{cr}$ to $\eta_\textrm{cr}$, gives 
\begin{eqnarray}
& & \frac{t}{t_0}=\frac{1}{6}\mbox{sinh}(2\eta)-\frac{2}{3}\eta+\frac{\sqrt{3}}{6}-\frac{2}{3}\eta_\textrm{cr},\,\,\,\eta\le-\eta_\textrm{cr}, \nonumber \\
& & \frac{t}{t_0}=\frac{1}{6}\mbox{sinh}(2\eta)-\frac{2}{3}\eta-\frac{\sqrt{3}}{6}+\frac{2}{3}\eta_\textrm{cr},\,\,\,\eta\ge\eta_\textrm{cr}.
\label{param2}
\end{eqnarray}
The parametric Eqs. (\ref{param1}) and (\ref{param2}) determine $\beta(t)$.
Putting $\beta(t)$ in (\ref{int}), written by means of (\ref{minimum}) as
\begin{equation}
a=\frac{a_\textrm{r}\beta}{\beta_\textrm{r}}\mbox{exp}\biggl(\frac{\beta^2_\textrm{cr}}{2\beta^2}\biggr)=\frac{a_\textrm{cr}\beta}{\sqrt{e}\beta_\textrm{cr}}\mbox{exp}\biggl(\frac{\beta^2_\textrm{cr}}{2\beta^2}\biggr),
\label{dynamics2}
\end{equation}
gives the dynamics of the early Universe, $a(t)$.

\begin{figure}[th]
\centering
\includegraphics[width=4in]{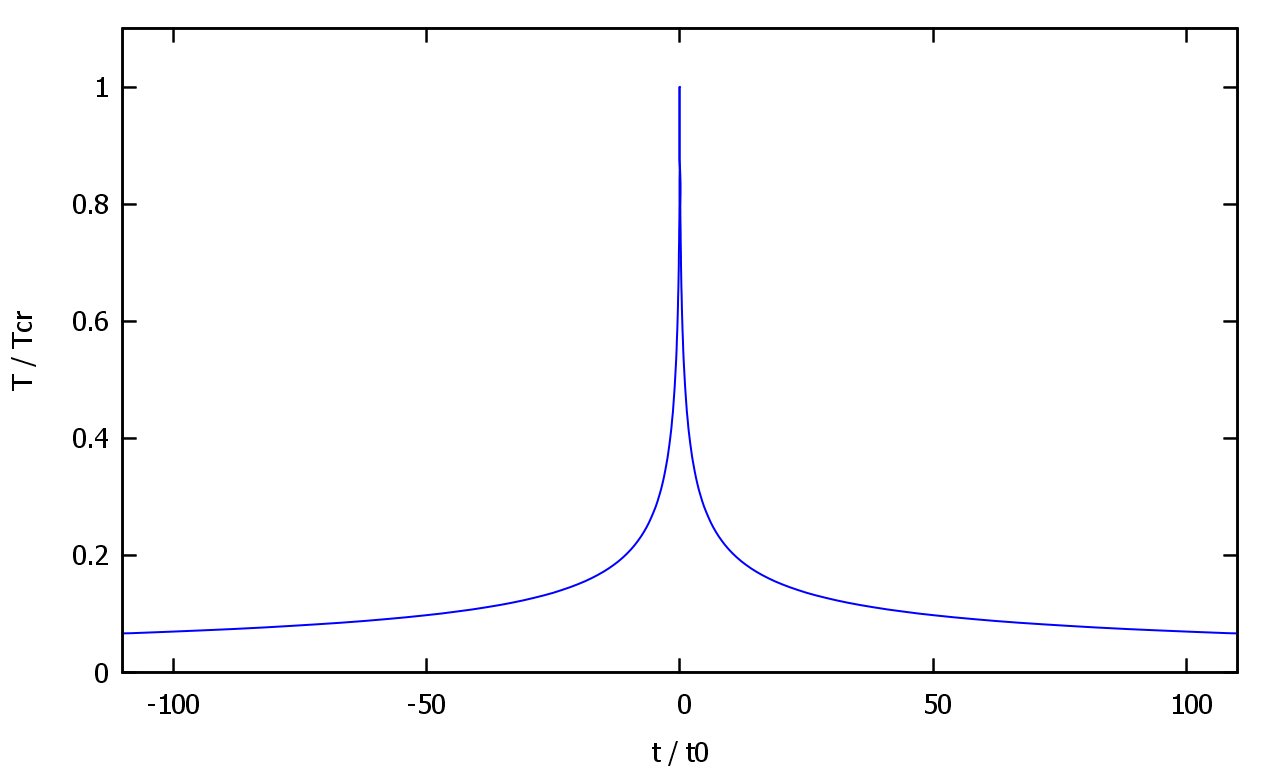}
\caption{The normalized temperature $\frac{T}{T_\textrm{cr}}$ as a function of the normalized cosmic time $\frac{t}{t_0}$.}
\label{fig_temp}
\end{figure}

\begin{figure}[th]
\centering
\includegraphics[width=4in]{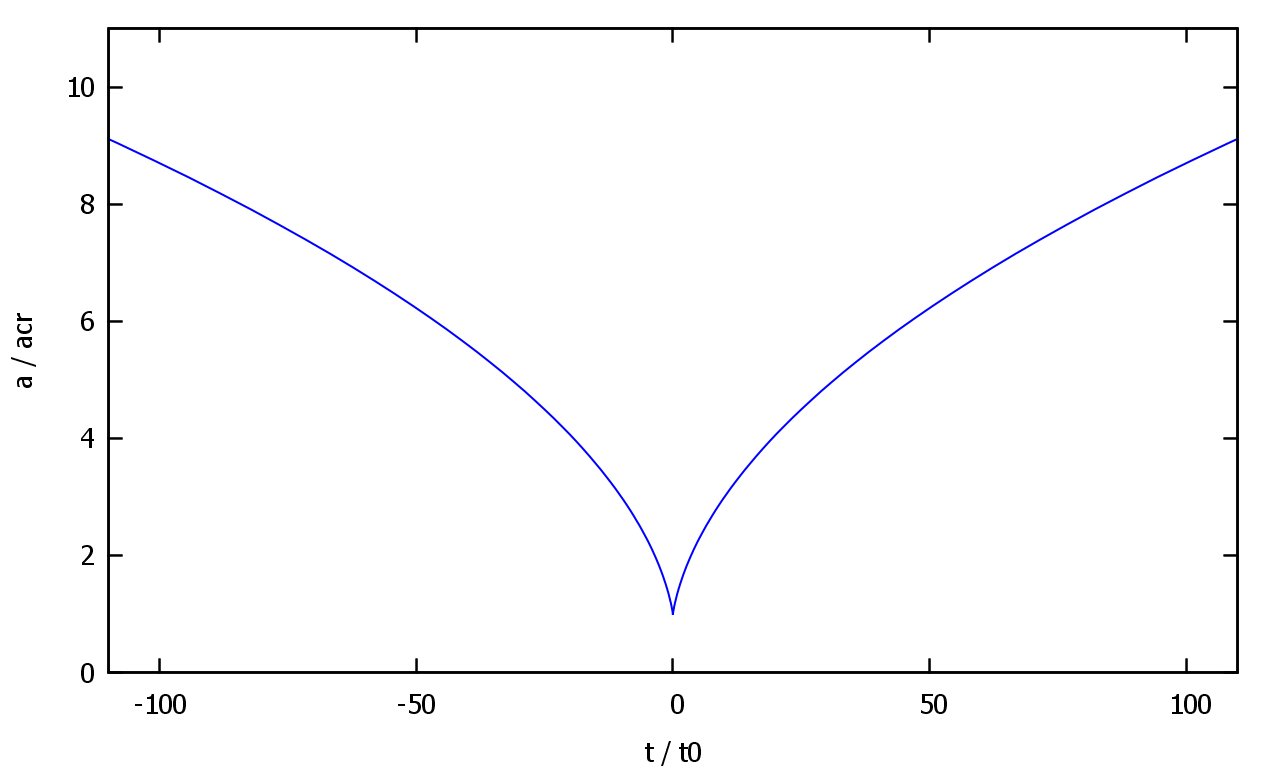}
\caption{The normalized scale factor $\frac{a}{a_\textrm{cr}}$ as a function of the normalized cosmic time $\frac{t}{t_0}$.}
\label{fig_scale}
\end{figure}

Figs. \ref{fig_temp} and \ref{fig_scale} show how the temperature and the scale factor, respectively, depend on the cosmic time.
The bouncing point, where the scale factor has its minimal value $a_\textrm{cr}$, is a cusp.
As $\eta$ increases from some initial, negative value, the Universe contracts ($\dot{a}<0$) until $\eta=-\eta_\textrm{cr}$, at which $\beta=\beta_\textrm{cr}$ and $a=a_\textrm{cr}$.
Then it undergoes a cusp-like bounce, at which $\eta$ jumps from $-\eta_\textrm{cr}$ to $\eta_\textrm{cr}$.
After the bounce, $\eta$ increases to infinity and the Universe expands ($\dot{a}>0$).
The unphysical big-bang singularity appearing in general-relativistic cosmology is replaced in the ECSK gravity by a nonsingular (with respect to curvature) big bounce that follows a contracting phase of the Universe \cite{Kuch}.
For $\beta\gg\beta_\textrm{cr}$, (\ref{dynamics1}) and (\ref{dynamics2}) give $a\propto T^{-1}\propto t^{1/2}$, which is characteristic to the radiation-dominated era.

The condition $\dot{a}=0$, defining the scale factor at a stationary state $a=a_\textrm{st}$, would be satisfied at a temperature $T_\textrm{st}$ given by (\ref{Fri1}) without the negligible term $k=1$: $h_\star T^4-\alpha h^2_n T^6=0$.
This temperature is equal to
\begin{equation}
T_\textrm{st}=\biggl(\frac{h_\star}{\alpha h^2_n}\biggr)^{1/2}>T_\textrm{cr},
\label{stationary}
\end{equation}
so the Universe never reaches $T=T_\textrm{st}$ and $\dot{a}=0$.
At the minimum scale factor $a_\textrm{cr}$, the Universe undergoes a cusp-like bounce from $\dot{a}=-v$ to $\dot{a}=v$, where
\begin{equation}
v=|\dot{a}(T_\textrm{cr})|=\biggl(\frac{\kappa}{3}(h_\star T^4_\textrm{cr}-\alpha h^2_n T^6_\textrm{cr})a^2_\textrm{cr}\biggr)^{1/2}=\biggl(\frac{32e}{243}\biggr)^{1/2}\frac{h_\star}{h_n}a_\textrm{r} T_\textrm{r}.
\label{velocity}
\end{equation}
At the bounce, the velocity of the point that is antipodal to the coordinate origin in a closed Universe is equal to $v_\textrm{ant}(T_\textrm{cr})=\pi v$ \cite{infl}.
The density parameter at the bounce is given by \cite{infl}
\begin{equation}
\Omega(T_\textrm{cr})=1+\frac{1}{v^2}=1+\frac{243 h^2_n}{32e h^2_\star(a_\textrm{r} T_\textrm{r})^2}.
\label{density}
\end{equation}
As the Universe expands, the antipodal velocity decreases according to
\begin{equation}
v_\textrm{ant}(T)=\pi\dot{a}(T)=\pi\biggl(\frac{\kappa}{3}(h_\star T^4-\alpha h^2_n T^6)a^2\biggr)^{1/2}=\pi\biggl(\frac{\kappa}{3}(h_\star T^2-\alpha h^2_n T^4)\biggr)^{1/2}a_\textrm{r}T_\textrm{r}\mbox{exp}\biggl(\frac{T^2}{2T^2_\textrm{cr}}\biggr),
\label{speed}
\end{equation}
and the density parameter increases according to
\begin{equation}
\Omega(T)=1+\frac{1}{\dot{a}^2(T)}=1+(a_\textrm{r} T_\textrm{r})^{-2}\biggl(\frac{\kappa}{3}(h_\star T^2-\alpha h^2_n T^4)\biggr)^{-1}\mbox{exp}\biggl(-\frac{T^2}{T^2_\textrm{cr}}\biggr).
\label{parameter}
\end{equation}
When the Universe reaches the radiation-dominated era, $T\ll T_\textrm{cr}$, (\ref{speed}) reduces to $v_\textrm{ant}(T)\sim T$ and (\ref{parameter}) reduces to $\Omega(T)-1\sim T^{-2}$.

Kerlick has shown that Dirac fields in the ECSK gravity satisfy the strong energy condition for the Hawking-Penrose singularity theorems \cite{Wald}, and that the torsion-induced spin-spin interaction (the second term on the right of (\ref{combined})) enhances that condition \cite{Ker}.
Such an interaction is thus attractive.
O'Connell has obtained the same result using the spin contributions to the Lagrangian density \cite{OC}.
Writing $\sigma_{ik}=T_{ik}+U_{ik}$, the strong energy condition holds if $W=(\sigma_{ik}-\frac{1}{2}g_{ik}\sigma^l_{\phantom{l}l})u^i u^k\ge0$.
For a homogeneous Universe, in the comoving frame of reference of an irrotational cosmological fluid, the tensor (\ref{combined}) and the Dirac equation give $W=\frac{1}{2}m\bar{\psi}\psi+\frac{3}{2}\kappa{\bf s}^2$, suggesting that torsion enhances the formation of singularities \cite{Ker}.
This result agrees with the contribution to $W$ from the torsion-induced spin-spin interaction, $\tilde{\epsilon}+3\tilde{p}=2\alpha n^2>0$.
The singularity theorems, however, have another important component: the Raychaudhuri equation, which describes the time evolution of the expansion scalar for timelike congruences \cite{Ray}.
The expansion scalar $\theta=u^i_{\phantom{i}:i}$ measures the fractional rate at which a small volume of matter changes with time as measured by a comoving observer.
Because of the Raychaudhuri equation, timelike geodesics for the matter satisfying the strong energy condition converge ($\theta$ decreases), resulting in a caustic, $\theta\rightarrow -\infty$ (if $\theta$ decreases continuously), and thus in a singularity within a finite proper time \cite{Ray}.

For the cosmological fluid in the comoving frame, the FLRW metric gives $\theta=(\mbox{ln}\sqrt{-g})_{,0}=\frac{3\dot{a}}{a}$.
Since the Universe at the minimum scale factor $a_\textrm{cr}$ undergoes a bounce from $\dot{a}=-v$ to $\dot{a}=v$, the expansion scalar has a discontinuity there, increasing from $\theta=-\frac{3v}{a_\textrm{cr}}$ to $\theta=\frac{3v}{a_\textrm{cr}}$.
As the Universe before the big bounce contracts, $\theta$ decreases from some value given by the initial condition of the contraction to a value $-\frac{3v}{a_\textrm{cr}}$.
At $a=a_\textrm{cr}$, $\theta$ jumps to $\frac{3v}{a_\textrm{cr}}$.
As the Universe after the bounce expands, $\theta$ decreases towards zero.
The discontinuity of the expansion scalar at the cusp-like big bounce therefore prevents $\theta$ from decreasing to $-\infty$ and reaching a curvature singularity, and guarantees that timelike geodesics in the Universe with Dirac fields coupled to torsion continue through the bounce.
A similar discontinuity completes timelike geodesics in the gravitational field of an Einstein-Rosen bridge \cite{radial}.

If we assume that the early Universe contained only known standard-model particles, then $g_\textrm{b}=28$ and $g_\textrm{f}=90$ \cite{cosmo}, so the temperature at the big bounce (\ref{critical}) was
\begin{equation}
T_\textrm{cr}\approx 0.78\,m_{\textrm{P}}.
\label{temperature}
\end{equation}
As the reference values, we can take the temperature and scale factor at the matter-radiation equality, where the radiation-dominated era ends.
We therefore have $T_\textrm{r}=T_\textrm{eq}\approx 0.75$ eV and $a_\textrm{r}=a_\textrm{eq}=\frac{a_0}{1+z_\textrm{eq}}$, where $a_0\approx 2.9\times10^{27}$ m is the present scale factor \cite{infl} and $z_\textrm{eq}\approx 3200$ is the redshift at the matter-radiation equality \cite{cosmo}.
Accordingly, the values at the big bounce of the scale factor (\ref{minimum}), antipodal velocity and density parameter were, respectively:
\begin{eqnarray}
& & a_\textrm{cr}\approx 5.9\times10^{-4}\,\mbox{m}, \label{scalefactor} \\
& & v_\textrm{ant}(T_\textrm{cr})\approx 8.9\times 10^{34}, \label{antipodal} \\
& & \Omega(T_\textrm{cr})\approx 1+1.3\times 10^{-70}.
\label{finetuning}
\end{eqnarray}
The value (\ref{antipodal}) is enormous, whereas (\ref{finetuning}) differs from 1 by a quantity that is extremely small in magnitude, like in the spin-fluid cosmology with torsion \cite{infl}.
Such extreme values result from $a_\textrm{r} T_\textrm{r}\gg 1$.

The big-bounce value of $\Delta=\Omega-1$ is on the order of $\frac{\Omega_S\Delta}{\Omega^2_R}\bigl|_0$, where $\Omega_S$ is the spinor-torsion density parameter \cite{infl}, subscript $R$ denotes radiation, and 0 denotes the present value.
An enormous value of $a_\textrm{r} T_\textrm{r}$ is related to an extremely small magnitude of $\Omega_S$:
\begin{equation}
\frac{\Omega_S\Delta}{\Omega^2_R}\biggl|_0\sim\frac{(G{\bf s}^2/\rho_c)\Delta}{(\epsilon/\rho_c)^2}\biggl|_0\sim\frac{Gn^2\rho_c\Delta}{\epsilon^2}\biggl|_0\sim\frac{GT^6\rho_c\Delta}{T^8}\biggl|_0\sim\frac{H^2\Delta}{T^2}\Bigl|_0\sim(aT)^{-2}\Bigl|_0\approx(a_r T_r)^{-2},
\label{scaling_infl}
\end{equation}
where $\rho_c$ is the critical density and $H$ is the Hubble parameter.
The apparent fine tuning of $\Omega(T_\textrm{cr})$ is thus caused by $|\Omega_S|\ll 1$, originating from an extremely weak spinor-torsion coupling in the ECKS gravity, as in \cite{infl}.
Accordingly, such a coupling naturally explains why the present Universe at largest scales appears nearly flat, solving the flatness problem without introducing exotic matter fields necessary for cosmological inflation.

This coupling is also responsible for an extremely rapid expansion of the Universe after the big bounce.
Such an expansion produces an enormous number of causally disconnected volumes, $N\sim v^3_\textrm{ant}$, from a single causally connected region (the closed Universe before and at the big bounce), naturally explaining why the present Universe at largest scales appears homogeneous and isotropic.
The spinor-torsion coupling therefore solves the horizon problem without inflation.
The transition from the torsion-dominated era to the radiation-dominated era occurs naturally as the contribution from this coupling to the Friedman equations (\ref{Fri1}) and (\ref{Fri2}) rapidly weakens (according to $T^{-6}$), which is another advantage of this scenario.

\section*{Acknowledgements}
I am grateful to Chris Cox, Bo\.{z}enna Pop{\l}awska and Janusz Pop{\l}awski for their support.
I would like to thank Sergei Kopeikin for the hospitality and fruitful discussions at the University of Missouri in Columbia.
I would also like to thank James Bjorken, Maurizio Gasperini, Richard Hammond, Friedrich Hehl, David Kerlick, Tom Kibble, and Robert O'Connell for valuable correspondence regarding torsion.
In addition, I would like to thank the referee for helpful comments which improved this paper.

\end{document}